\documentclass[pdflatex,sn-basic]{sn-jnl}

\usepackage{graphicx}%
\usepackage{multirow}%
\usepackage{amsmath,amssymb,amsfonts}%
\usepackage{amsthm}%
\usepackage{mathrsfs}%
\usepackage[title]{appendix}%
\usepackage{xcolor}%
\usepackage{textcomp}%
\usepackage{manyfoot}%
\usepackage{booktabs}%
\usepackage{algorithm}%
\usepackage{algorithmicx}%
\usepackage{algpseudocode}%
\usepackage{listings}%
\usepackage{url}
\usepackage{longtable}
\usepackage{subcaption}
\usepackage{array}
\usepackage{float}
\usepackage{makecell}

\theoremstyle{thmstyleone}%
\theoremstyle{thmstyletwo}%

\theoremstyle{thmstylethree}%

\begin{document}

\title[Article Title]{AI-Enabled grading with near-domain data for scaling feedback with human-level accuracy}


\author[1]{\fnm{Shyam} \sur{Agarwal}\footnotemark[0]}\email{shyagarwal@ucdavis.edu}

\author*[2]{\fnm{Ali} \sur{Moghimi}\footnotemark[0]}\email{amoghimi@ucdavis.edu}

\author[3]{\fnm{Kevin C.} \sur{Haudek}\footnotemark[0]}\email{haudekke@msu.edu}

\footnotetext[1]{ORCID: \href{https://orcid.org/0009-0009-2147-5674}{0009-0009-2147-5674}}
\footnotetext[2]{ORCID: \href{https://orcid.org/0000-0002-9249-8365}{0000-0002-9249-8365}}
\footnotetext[3]{ORCID: \href{https://orcid.org/0000-0003-1422-6038}{0000-0003-1422-6038}}

\affil[1]{\orgdiv{Department of Computer Science}, \orgname{University of California, Davis}, \orgaddress{\street{One Shields Ave}, \city{Davis}, \postcode{95616}, \state{CA}, \country{USA}}}

\affil[2]{\orgdiv{Department of Biological and Agricultural Engineering}, \orgname{University of California, Davis}, \orgaddress{\street{One Shields Ave}, \city{Davis}, \postcode{95616}, \state{CA}, \country{USA}}}

\affil[3]{\orgdiv{Department of Biochemistry \& Molecular Biology and CREATE for STEM Institute}, \orgname{Michigan State University}, \orgaddress{ \city{East Lansing}, \postcode{48824}, \state{Michigan}, \country{USA}}}


\abstract{Constructed-response questions are crucial to encourage generative processing and test a learner’s understanding of core concepts. However, the limited availability of instructor time, large class sizes, and other resource constraints pose significant challenges in providing timely and detailed evaluation, which is crucial for a holistic educational experience. In addition, providing timely and frequent assessments is challenging since manual grading is labor intensive, and automated grading is complex to generalize to every possible response scenario. This paper proposes a novel and practical approach to grade short-answer constructed-response questions. We discuss why this problem is challenging, define the nature of questions on which our method works, and finally propose a framework that instructors can use to evaluate their students' open-responses, utilizing near-domain data like data from similar questions administered in previous years. The proposed method outperforms the state of the art machine learning models as well as non-fine-tuned large language models like GPT 3.5, GPT 4, and GPT 4o by a considerable margin of over 10-20\% in some cases, even after providing the LLMs with reference/model answers. Our framework does not require pre-written grading rubrics and is designed explicitly with practical classroom settings in mind. Our results also reveal exciting insights about learning from near-domain data, including what we term as accuracy and data advantages using human-labeled data, and we believe this is the first work to formalize the problem of automated short answer grading based on the near-domain data.}




\maketitle

\section{Introduction and Related Works}

Assessments play a significant role in a student's learning outcomes. On one hand, they provide valuable information to students about where they are lacking and what improvements they can make. On the other hand, they can inform instructors about student progress and understanding of the various concepts being taught. Well-designed assessments can also promote the development of students' logical aptitude, problem-solving, critical thinking, and other cognitive skills. Feedback from these assessments has a strong influence on their learning and achievement \citep{hattie2007power}. In fact, many historical works have identified the importance of feedback in either improving knowledge and acquiring skills \citep{azevedo1995meta, bangert1991instructional, corbett1989feedback, epstein2002immediate, moreno2004decreasing, pridemore1995control} or an essential factor in motivating the learning process \citep{lepper1985intrinsic, narciss2004design}.

With an increasing demand for quality education, class sizes have grown considerably, while the availability of instructors and teaching assistants to assess student learning remain limited. This creates significant time constraints, increasing the need for efficient learning support that can provide timely feedback to the learners. Scaling this feedback has been a widely accepted challenge, often referred to as the "feedback" challenge \citep{Wu2018ZeroSL}.

To solve these problems, there have been efforts to leverage technology, especially AI models to automate feedback generation, addressing the existing constraints in education. Automated grading has emerged as the first step in this process because once we know which category the response falls in (correct or incorrect, for example), then it becomes easier to provide template-based feedback or more personalized feedback with some additional overhead work.

It is important to note that the challenge in the classification or evaluation of responses largely depends on the type of question being considered. Objective questions, such as multiple-choice,  are relatively straightforward to grade since the possible response choices are limited. However, these questions do not provide much insight into a student's learning journey. In contrast, constructed-response questions (CRQ) provide a great way to understand student thinking more deeply, serving as a powerful means of evaluating a learner's understanding of concepts. These questions require learners to organize their thoughts, provide explanations, and demonstrate higher-level thinking skills which further requires an ability to reason through elaborate answers \citep{Zhao2021TargetedFG}. Short answers are a type of CRQ that asks learners to write their answers in a few sentences. The grading of these questions focuses on the content, instead of the writing quality. Despite their effectiveness, compared to objective questions, CRQs are significantly more challenging to grade due to the complexity and variability of student responses.

Although Automated Short Answer Grading (ASAG) is a widely studied problem, it has emerged to be a hard problem and the following challenges have been identified as the major causes for it and other human-centered AI tasks \citep{Wu2018ZeroSL, malik2021generative}:

\begin{enumerate}
\item Student work shows a great amount of diversity, with many responses being unique (long tail/zipf distribution). Thus, statistical supervised grading techniques would find it hard to produce promising results.
\item The manual labeling of responses is subjective, laborious, time-consuming and expensive. Thus, access to realtively large annotated datasets is difficult.
\item Grading is a domain where not only accuracy but also precision (reproducibility) is paramount due to the high stakes involved in misgrading.
\item The entire grading process must be transparent and justifiable to instructors and students alike.
\end{enumerate}

ASAG has a strong literature background with people using many approaches, from traditional statistical ones to modern day neural networks \citep{zhai2020applying}. The choice of what approach to use mostly boils down to the availability of enough manually annotated dataset. If enough data is available to train the models, one would likely assume that high performance can be obtained by just using supervised learning techniques or statistical models. This is in part due to the ability to fine tune models to learn patterns present in domain specific tasks, like evaluating science explanations or reasoning \citep{jescovitch2021comparison}.

Various methods have been studied for this application. For short-answer grading, \citet{heilman2013ets} took a domain adaptation approach with using word and character n-gram features. \citet{sultan2016fast} employs supervised learning methods like linear regression, and \citet{hou2011automatic} uses support vector machines and incorporates POS tags and term frequency to get good results. \citet{madnani2013automated} built a logistic regression classifier using simple features like count of commonly used words and the length of responses. 
 
 The performance of such conventional ML models won’t necessarily increase as we feed more data as they eventually get to a plateau and saturate. Even in the case of achieving promising results, conventional machine learning models are often limited by their lack of generalization. These models are typically highly specialized and less capable of generalizing across new or unseen questions as they are trained on fixed sets of questions and corresponding answers. This results in poor performance when attempting to grade answers to questions that weren't part of the initial training set, even if they belong to the same domain. Additionally, these models require significant effort to update or retrain to accommodate new data or question formats.

Many researchers have also used embedding-based neural networks. \citet{riordan2017investigating} used standard neural architectures using conventional word embeddings, attention, and LSTM layers on a number of varied datasets. The list goes on with \citet{zhang2016deep} using a combination of deep belief networks and feature engineering, \citet{yang2018automatic} using deep autoencoder models, \citet{liu2019automatic} using multi-way attention networks, \citet{tan2020automatic} encoding student responses using Graph Convolutional Networks and \citet{qi2019attention} using CNN and Bi-LSTM models to create hierarchical word-sentence model. Transformer architectures have also been widely used by a number of researchers. \citet{sung2019pretraining} achieved human-comparable performance by just fine-tuning BERT model on a domain-specific task. \citet{gaddipati2020comparative} evaluated different response embeddings for their ASAG task, namely ELMo, GPT, BERT, and GPT-2. The size of transformers and their ability to generalize to different languages has also been studied \citep{camus2020investigating}. However, the availability of large amounts of data is a concern here. In most practical settings, it is rare to find such large amounts of training data. Instead, most instructors have access to near-domain data (data pertaining to the same overarching topic), which is what we leverage in our work.

Most recently, people have started using large language models for these grading tasks \citep{henkel2023can, schneider2024towards}. Although LLMs are not trained for classification tasks, many works have shown promising results on these tasks by using few-shot prompting techniques giving few examples in the system prompt \citep{cohn2024chain, kortemeyer2023performance, ouyang2022training}. \citet{golchin2024grading} attempted to replace peer grading in massive open online courses using OpenAI's GPT 3.5 and GPT 4 models. \citet{yoon2023short} uses one-shot prompting and text similarity scoring (aligned with the three-step approach). \citet{kortemeyer2023performance} utilized the standard benchmark 2-way and 3-way datasets SciEntsBank and Beetle, and studied the performance of OpenAI's GPT-4 on them with and without a reference answer.

Keeping the above approaches in mind and the recent progresses in large language models like BERT and GPT, we show that the necessary algorithms to achieve the state-of-the-art performance for evaluating science assessments already exist if utilized in a proper manner. The rest of this paper discusses this approach and shares our insights gathered throughout the process. Our main contributions can be summarized as following:

\begin{enumerate}
\itemsep=0pt
\item We provide a novel framework to automate the grading of short constructed-response questions that requires a very limited amount of data to achieve a competitive performance, especially because it relies on near-domain data like similar questions administered in previous years.
\item We compare the performance of high-compute (large number of parameters) models like GPT 3.5 Turbo, GPT 4 Turbo, and GPT 4o with the performance of low-compute (smaller number of parameters) models like BERT (and its fine-tuned versions).
\item We identify and analyze the patterns of mistakes that the models make.
\item We provide a discussion on the trade-offs between model size, environmental impact, and how the framework is applicable in practical settings, ensuring consistency, fairness, and efficiency in automated grading systems.
\end{enumerate}

\section{Methodology}
As shared earlier, we propose a novel method to grade the short-answer responses, which is the first step in the feedback generation process.
\subsection{Dataset}

Our dataset consists of three related CRQs about the biological information flow (Central Dogma of Molecular Biology). In this section, we share some important insights about the dataset.
\subsubsection{Significance of the underlying topic}
The topic of information flow has been acknowledged as a fundamental biological concept by many researchers \citep{AAAS2011, NGSS2013}. However, both secondary and postsecondary level students have shown difficulty in understanding it \citep{Prevost2016}. As highlighted by \citet{Lewis2000} and \citet{MillsShaw2008}, the difficulty for many students lies in understanding how the genetic information is stored and exchanged. In contrast, many students find it difficult to understand how the genetic information (DNA) is encoded into proteins through a RNA intermediate. Many works have highlighted student misconceptions about these topics \citep{Fisher1985, MillsShaw2008, WoodRobinson2000, MarbachAd2001, SmithKnight2012, SmithWoodKnight2008}.

Many researchers \citep{WrightFiskNewman2014, Prevost2016} have tried to understand why this barrier in understanding the topic of central dogma exists, often focusing on root causes in pedagogy such as the use of uninsightful diagrams for explaining the concept. In contrast, this study aims to further previous work using an automated text classification system to assess students' understanding of the process and help instructors identify patterns of mistakes.

\subsubsection{Origin of the Dataset}
The Genetics Concept Assessment (GCA) is a concept inventory mainly focused on information flow, exchange and storage which are fundamental ideas in many undergraduate biology and genetics courses  \citep{SmithWoodKnight2008}. The instrument reveals student understanding of nine learning goals through twenty-five multiple choice questions. For this study, we employ a version of a GCA multiple-choice question that was modified by \citet{Prevost2016}. The modified item tests understanding of a core concept in undergraduate biology (i.e. information flow) \citep{aaas2010vision} and maintains alignment to the original learning goal in the GCA. This modified item is composed of three, related, open-ended prompts that asked students to explain the effect of a mutation in a strand of DNA that results in a premature stop codon.  Students are asked to explain the effect on three closely related processes in information flow, namely replication, transcription, and translation. The question is as follows:
\textit{The following DNA sequence occurs near the middle of the
coding region of a gene.
DNA 5’ A A T G A A T G G* G A G C C T G A A G G A 3’
There is a G to A base change at the position marked with
an asterisk. Consequently, a codon normally encoding an
amino acid becomes a stop codon.
\begin{itemize}
    \item \textbf{Q1:} How will this alteration influence DNA replication?  
    \item \textbf{Q2:} How will this alteration influence DNA transcription?  
    \item \textbf{Q3:} How will this alteration influence DNA translation?  
\end{itemize}}

Some sample responses for each of these questions along with their classification is shown in table \ref{table:sampleoutput}. These questions were administered to undergraduate biology in introductory cell and molecular biology-focused courses at multiple large public research universities in the US as a part of their regular online homework assignments after the relevant content on central dogma was taught. The completion of the assignment was rewarded with one point and students were encouraged to just try their best. A total of 2,861 students participated, each providing one response per question across the three questions. These responses were marked as either correct, incomplete or incorrect by human experts with an inter-grader agreement of 80\% \citep{Prevost2016, bierema2021quantifying}. Majority vote was accounted for when any disagreements happened. Some responses were removed from the data because they were nonsensical (like "IDK"). We believe that the data is representative since it is coming from multiple universities, class populations, and time periods. The data distribution is shown in table \ref{table:datadistribution}.

{\small
\begin{longtable}{l p{7cm} l}
\caption{Sample responses and model classification/evaluation for questions on the effects of a genetic alteration on Replication, Transcription, and Translation}
\label{table:sampleoutput} \\

\toprule
Question & Response & Classification \\
\midrule
\endfirsthead

\toprule
Question & Response & Classification \\
\midrule
\endhead

\midrule
\multicolumn{3}{r}{\small\itshape Continued on next page...} \\
\midrule
\endfoot

\bottomrule
\endlastfoot


\midrule

Effect on Replication & This will have a severe effect on DNA replication. Once the DNA polymerase reads the stop codon it stops replicating the nucleotides in the DNA code. This will leave a major chunk of the gene to be lost due to the fact that it is not replicated because the DNA polymerase reached a stop codon before it was supposed to. & Incorrect \\

\midrule

Effect on Replication & It wouldn't really change the speed of replication at all. The strand that had this mutation on it would be mutated, so the effect for that copy would be permanent, but there would be no evidence of mutation on the other strand, so it would still be a G in that daughter strand. In summary, you'd get one daughter with A and one with G. I don't believe the proofreading mechanisms on the polymerase would catch it since they only catch errors that they make, and this error happened at another time. & Correct \\

\midrule

Effect on Replication & A mutation that results in a stop codon in the middle of a DNA sequence results in the stopping of translation and therefore the protein is not formed. This changes DNA replication because not all of the protein is formed which messes up the DNA of the cell. & Incomplete \\

\midrule

Effect on Transcription & The stop codon signals the end of the DNA sequence that is to be transcribed. During transcription the DNA sequence is copied to RNA and when the G to A base change occurs a stop codon is produced resulting in that codon as the last to be transcribed. Because the mRNA ends with the stop codon only. & Incorrect \\

\midrule

Effect on Transcription & This alteration will cause a mutation in the DNA replication. A point mutation occurs and then DNA replication will begin replicating the mutated gene sequence. A stop code is a nucleotide triplet within messenger RNA that signals a translation. The stop codon tells the ribosomes executing the mRNA code that it's time to stop. & Incomplete \\

\midrule

Effect on Transcription & Because DNA is always replicated and transcribed from 5' to 3', mRNA will transcribe the chain into UUA/CUU/ACC, with ACC acting as the stop codon. Because there is a stop codon, the following nucleotides will not be translated into amino acids. However, mRNA will continue to transcribe the DNA into mRNA strands of UUA/CUU/ACC/CUC/GGA/CUU/CCU, but they will only be translated up to the stop codon. & Correct \\

\midrule

Effect on Translation & The stop codons are UAA, UAG, and UGA. None of these are in this strand of DNA even with the base change so nothing will happen with the stop codon. Translation will decode the mRNA that is made by replicating the template strand which will then produce polypeptides. & Incorrect \\

\midrule

Effect on Translation & The term point mutation originated before the advent of DNA sequencing and therefore before it was routinely possible to discover the molecular basis for a mutational event. Nowadays, point mutations typically refer to alterations of single base pairs of DNA or to a small number of adjacent base pairs. Examples include deletions, insertions, substitutions, etc. & Incomplete \\

\midrule

Effect on Translation & The alteration will be slightly detrimental in translation because the polypeptide chain will be very short and may not consist of all the amino acids needed to produce the right proteins needed by the body. In this sense then the body becomes deprived of some proteins and different diseases associated with this deprivation occur as a result. & Correct \\

\bottomrule

\end{longtable}
}

\begin{table}[h]
\caption{Distribution of collected data for Replication, Transcription, and Translation questions across training, validation, and testing sets, categorized as Correct, Incomplete, and Incorrect}\label{table:datadistribution}%
\begin{tabular}{@{}llllll@{}}
\toprule
Question & Label & Training & Validation & Testing & Total \\
\midrule
\multirow{3}{*}{Replication} & Correct & 785 & 219 & 554 & 1558 \\
                              & Incomplete & 240 & 64  & 169 & 473 \\
                              & Incorrect  & 405 & 146  & 279 & 830 \\
\midrule
\multirow{3}{*}{Transcription} & Correct & 759 & 217 & 554 & 1530 \\
                                 & Incomplete & 220 & 79 & 146 & 445 \\
                                 & Incorrect  & 451  & 133 & 302 & 886 \\
\midrule
\multirow{3}{*}{Translation} & Correct & 940 & 272 & 665 & 1877 \\
                               & Incomplete & 310  & 96  & 203 & 609 \\
                               & Incorrect  & 180  & 61 & 134 & 375\\
\midrule
\end{tabular}
\end{table}

\subsection{Terminology and Approach}
\label{sec:terminology-and-approach}
\subsubsection{Splitting Data}
We will use the following symbols to denote different aspects in our process. As described above, our dataset is composed of student responses to three different questions. We denote each student as $S_1$, $S_2$, $…$, $S_n$. $Q_1$ refers to the first question which deals with the effect on replication, $Q_2$ refers to the second question which deals with the effect on transcription, and $Q_3$ refers to the third question which deals with the effect on translation. The responses from a given student $S_k$ are written as $S_{k|1}$, $S_{k|2}$, $S_{k|3}$. From the data that we have, we reserve responses from 50\% (1430) of the students for training ($T_1$, $T_2$, $T_3$ for $Q_1$, $Q_2$, and $Q_3$ respectively) and 15\% (429) of the students for validation purposes ($V_1$, $V_2$, $V_3$ for $Q_1$, $Q_2$, and $Q_3$ respectively). The training data is used to train the model and adjust its parameters, while the validation data is used to evaluate the model’s performance, aiding in model tuning and preventing overfitting.

We ensure that all three responses from a single student remain within the same dataset partition. This partitioning guarantees that a student’s response to one question is never used to predict any characteristics of their responses to other questions. In other words, since we have responses to all three questions from a particular student, we do not train a model using some of their responses while testing it on others. This helps maintain generality throughout the process by ensuring that the model does not inadvertently rely on the dependency between responses from the same student across different questions, thereby making the results applicable to other experiments where responses from the same student to all three questions are considered independently.

The responses from the remaining 35\% (1002 samples) of students for $Q_1$, $Q_2$, and $Q_3$ are reserved to test generalization of the model on unseen datasets, hereby referred as $T'_1$, $T'_2$, $T'_3$ respectively. We have $S_1$, $S_2$, $…$, $S_n$ students in some random order and let $p = 0.5n$ and $q = 0.15n$, then for i = 1, 2, 3 for question $Q_i$, we have

\begin{align}
T_i &= \{S_{1|i}, S_{2|i}, \dots, S_{p|i}\} \\
V_i &= \{S_{p+1|i}, S_{p+2|i}, \dots, S_{p+q|i}\} \\
T'_i &= \{S_{p+q+1|i}, S_{p+q+2|i}, \dots, S_{n|i}\}
\end{align}

\subsubsection{Fine-tuning BERT models}
We start with a basic BERT model (hereby, referred as base model $B_0$), and fine tune it on all of $T_1$ to get $BMQ_1$, fine tune it on all of $T_2$ to get $BMQ_2$, and fine tune it on all of $T_3$ to get $BMQ_3$. $BMQ_1$ was evaluated on $T'_1$, $BMQ_2$ was evaluated on $T'_2$, and $BMQ_3$ was evaluated on $T'_3$. We did not evaluate these models on other questions because each model was fine-tuned specifically to categorize responses for one question only, and lacks the context to handle responses from other questions. Corresponding validation sets $V_1$, $V_2$, and $V_3$ were used during training for early stopping and model selection. Specifically, during training, if the validation accuracy did not improve for 10 consecutive epochs, training was stopped early and the model weights from the best-performing epoch were restored.

\subsubsection{Near-domain Data}
In our work, the questions share a common subject matter (information flow through the Central Dogma of molecular biology), which means that in order to predict classification scores for a particular question, the model trained on another question can also be leveraged. The data on which this other model is trained is what we refer to as near-domain data throughout the work. We wanted to confirm if a model trained on near domain data would require less training dataset to perform similarly to the model that is trained on a larger dataset, but without near domain data.

\subsubsection{Models leveraging near-domain data}
\begin{figure}
    \centering
	\includegraphics[width=0.7\columnwidth]{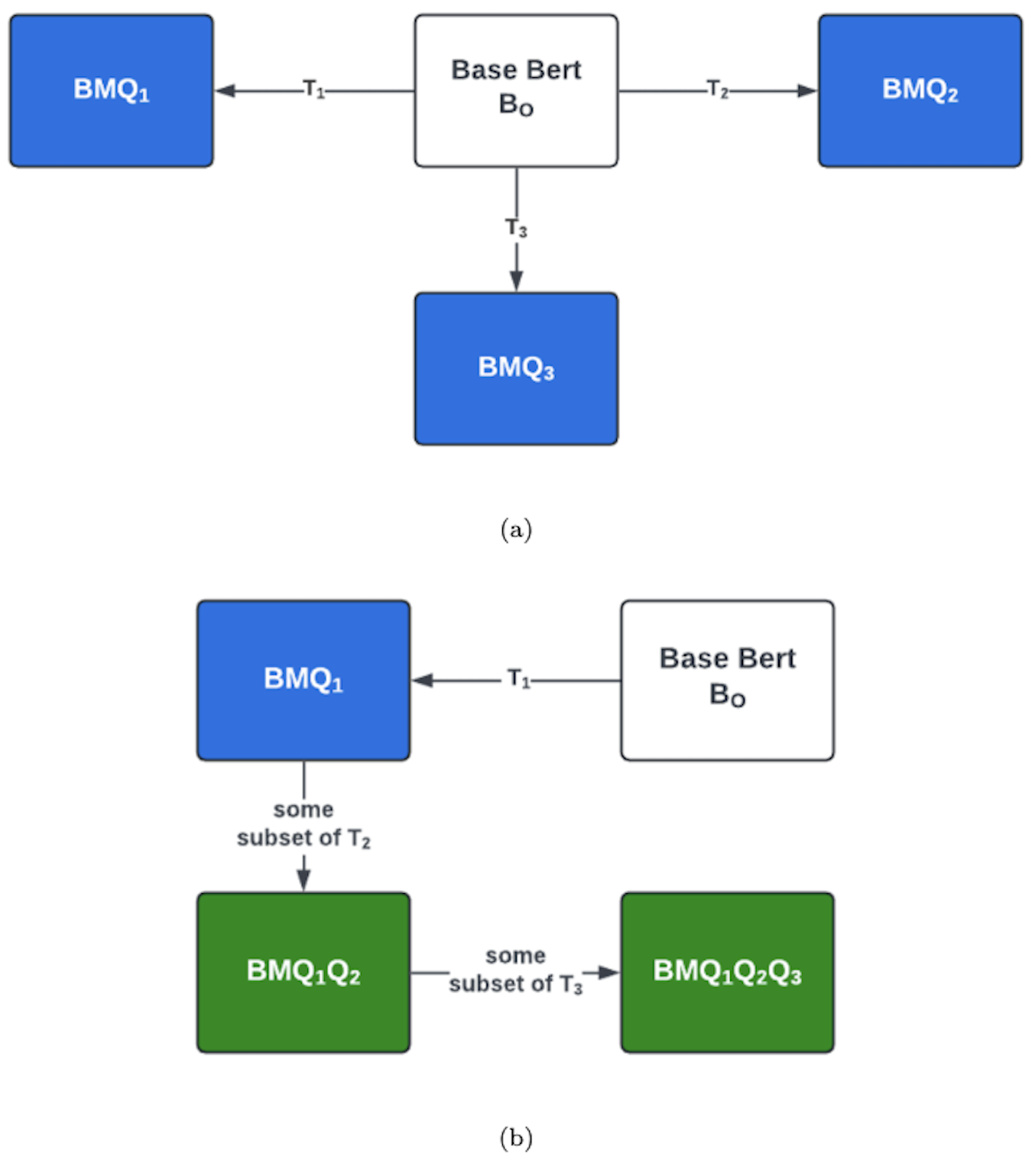}
	\caption{Model terminology and fine-tuning process: (a) $B_0$ represents the base BERT model, fine-tuned separately on training data $T_1$, $T_2$, $T_3$ to create $BMQ_1$, $BMQ_2$, $BMQ_3$ and (b) Near-domain fine-tuning is applied sequentially where $B_0$ is fine-tuned on $T_1$ to produce $BMQ_1$, $BMQ_1$ is fine-tuned on a subset of $T_2$ to produce $BMQ_1Q_2$, which is further fine-tuned on a subset of $T_3$ to obtain $BMQ_1Q_2Q_3$.}
    \label{FIG:1}
\end{figure}

In order to test how well a model can leverage the knowledge from one question to potentially improve the performance of others within the same domain, we fine-tune $BMQ_1$ on different subsets of $T_2$ to obtain the best model $BMQ_1Q_2$, and then we fine tune $BMQ_1Q_2$ on different subsets of $T_3$ to obtain the best model, $BMQ_1Q_2Q_3$. For reference, refer to Figure \ref{FIG:1} for an easier understanding of the terminology. The subsets of training data used for $BMQ_1...Q_j$ (j = 2, 3) models were chosen by using a fixed random seed and stratified sampling of the particular training dataset in increments of 2.5\% or about 35 student samples (from 0\% to 100\%). Throughout the training process, we use a batch size of 16 and drop out any remaining samples that could not be accounted in the batch. Again, as explained above, since any of these BERT and BERT derivative models do not have the context of the question, it does not make sense to evaluate them on responses to other questions. Thus, $BMQ_1$ was evaluated on $T'_1$, $BMQ_1Q_2$ was evaluated on $T'_2$ and $BMQ_1Q_2Q_3$ was evaluated on $T'_3$ only. The success of this strategy would imply that in the future, we would only need a small set of annotated samples for new questions, significantly reducing the amount of labeled data required to achieve optimal model performance. Annotating data for machine learning models, especially in specialized domains like this, is time-consuming and costly. Thus, if successful, this approach would allow us to leverage existing models trained on related questions, minimizing the need for extensive annotation efforts for each new question. This strategy, if proven effective, could have widespread implications for educational models and other domains where labeled data is scarce. Further details about the process are explained in \ref{sub:2.3.1}.

To identify the best-performing model for each question, we examined two factors - the evaluation metric (chosen as accuracy in this study) and the amount of training data used for fine-tuning. Specifically, for each $BMQ_1...Q_j$ (j = 2, 3) model, we assessed how accuracy changed as the percentage of training data from $T_j$ increased, helping us determine the most efficient or smallest subset size required for relatively optimal performance. To confirm the consistency and reliability of the accuracy metric obtained on the training data, we used the respective validation sets: $BMQ_1$ was evaluated on $V_1$, $BMQ_2$ and $BMQ_1Q_2$ were evaluated on $V_2$, and $BMQ_3$ and $BMQ_1Q_2Q_3$ were evaluated on $V_3$. On the testing data, we find the top 5 most accurate models, calculate their average accuracy, and choose the most accurate model with the least amount of data that is within 1 standard deviation of this average accuracy. Details of this can be found in table \ref{table:chosenmodels}.

\subsection{Model Architecture and Training}
Since the advent of BERT \citep{devlin2018bert} and GPT \citep{radford2018improving}, the dominant adaptation technique has been model tuning or fine-tuning \citep{lester-etal-2021-power} even for variants of GPT that came afterwards like \citep{radford2019language}. The two most common ways of fine-tuning include freezing the weights in the earlier layers and only changing the ones in the later layers or retraining all the parameters in the model as proposed by \citet{howard-ruder-2018-universal}.  We tried our experiments with both the approaches and found out that the latter produced results that were suboptimal by a huge margin of 10-20\% across all questions. Thus, the results shared in the rest of the paper are based on training all the parameters. We share the specific details around fine-tuning BERT model and OpenAI’s GPT models below:

\subsubsection{BERT}
\label{sub:2.3.1}
BERT (Bidirectional Encoder Representations from Transformers) is a language representation model based on the Transformer model \citep{vaswani2017attention} that is trained on a large corpus of unlabeled text. One particularly distinguishing characteristic of BERT is the Transformer model's self-attention mechanism that compares the significance of every word in a piece of sentence to every other word in that sentence enabling it to collect context in both the forward and the backward direction. This bidirectional nature helps BERT capture and represent deeper and richer language representations than traditional unidirectional models.\newline

The $B_0$ model consists of input IDs and attention masks mapping to the main layer, which comprises 12 layers and 12 attention heads with the pooler output having 768 nodes (totaling, 108,310,272 parameters). For $BMQ_1$, $BMQ_2$, and $BMQ_3$, we attach this main layer to an intermediate layer ($IL_1$) with 512 nodes (393,728 parameters) which is attached to a classification head ($CH_1$) of 3 nodes (1,539 parameters), each of which is associated with a single class (correct, incomplete, or incorrect class). All the parameters are trainable. 

For training these models, we employ the Adam Optimizer \citep{kingma2014adam} with a learning rate of $1e-5$, and use categorical cross entropy as the loss function. The number of epochs are guided by an early stopping callback function that monitors the validation accuracy for 10 successive epochs without an improvement before it restores the weights of the model performing the best on validation data.

\begin{figure}
	\centering
	\includegraphics[width=\columnwidth]{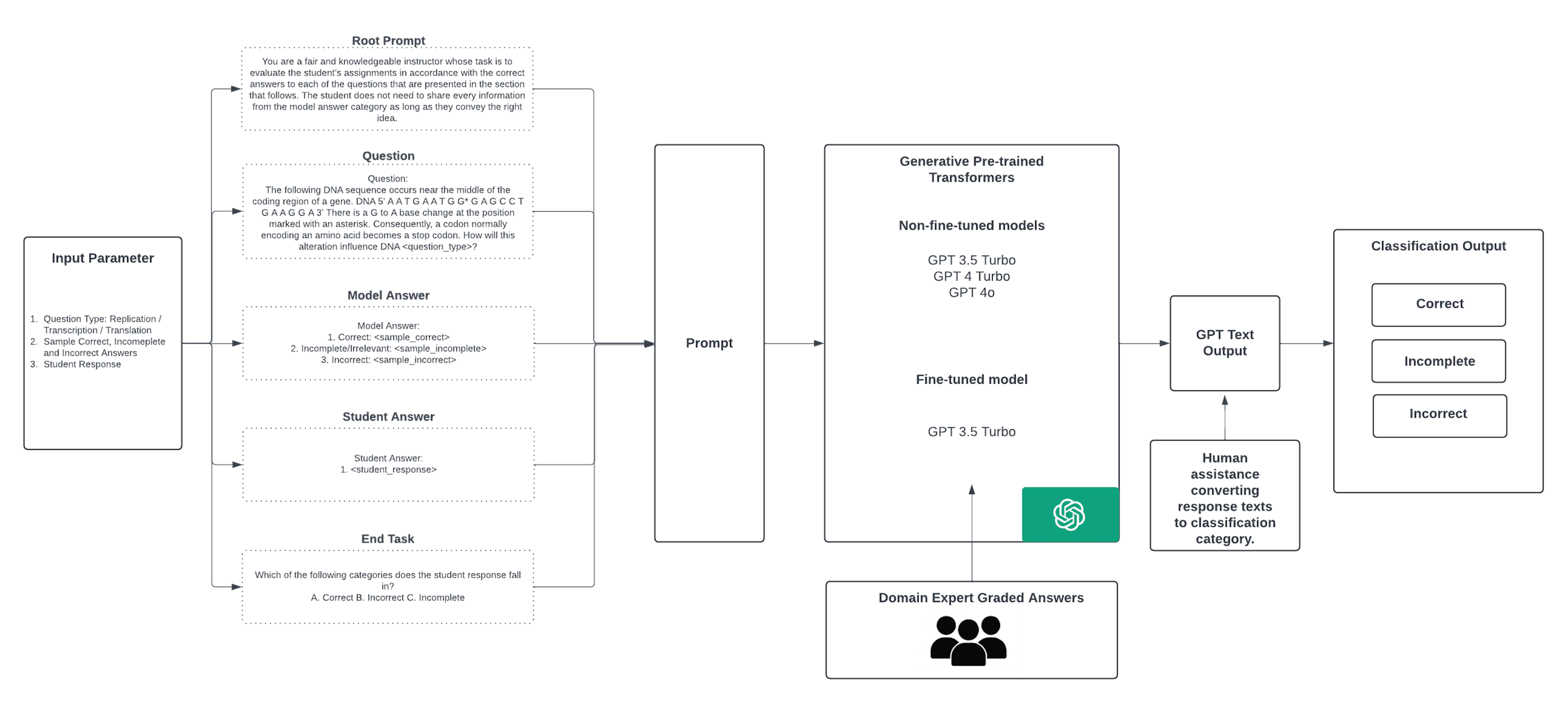}
	\caption{GPT-based student response classification pipeline:The question type (Replication, Transcription, or Translation), sample correct, incomplete, and incorrect answers, and the student response are added for different cases into the prompt which is then fed into pre-trained GPT models (both fine-tuned and non-fine-tuned versions) to generate classification outputs with Human assistance needed to map the model-generated text to classification categories (Correct, Incomplete, or Incorrect). Domain expert-graded answers serve as the ground truth for evaluation.}
	\label{FIG:2}
\end{figure}

\subsubsection{GPT Series}
\label{sub:2.3.2}
To understand how good LLMs are at grading these responses, we also perform several experiments with fine-tuned and non-fine-tuned models from OpenAI's GPT series. We directly prompt GPT 3.5 Turbo, GPT 4, and GPT 4o without fine-tuning.  We also fine-tune GPT 3.5 Turbo (the only model allowed to be fine-tuned at the time of the experiments), but for this fine-tuning, instead of using increments of 2.5\% to test the model performance, we used the entire training data. All these experiments are done using different temperatures (0, 0.5, and 1).

For GPT fine-tuning, we follow the standard procedure as suggested by OpenAI. For the prompt that is used to train the GPT models, we take inspiration from \citet{golchin2024grading} to decide our root prompt which is as follows: \newline \textit{You are a fair and knowledgeable instructor whose task is to evaluate the student’s assignments in accordance with the correct answers to each of the questions that are presented in the section that follows.\newline The student does not need to share every information from the model answer category as long as they convey the right idea.}

Then, we provide the question, the model answer, the student response, and ask the model to classify it as either correct, incorrect, or incomplete as a multiple-choice question as suggested and done by \citet{Santurkar2023Whose}. The rest of the prompt is given below, where $<$substring$>$ represents the answer provided by the student and $<$sample\_correct$>$, $<$sample\_incomplete$>$, and $<$sample\_incorrect$>$ refer to the sample correct, sample incomplete, and sample incorrect responses for each $<$question\_type$>$ (replication, transcription, translation). Table \ref{table:sampleresponse} shares the exact samples, which is inspired from \citep{Prevost2016}. Details about the entire pipeline are shown in Figure \ref{FIG:2}. \newline\newline \textit{
Question:\newline The following DNA sequence occurs near the middle of the coding region of a gene. \newline DNA 5' A A T G A A T G G* G A G C C T G A A G G A 3’ \newline There is a G to A base change at the position marked with an asterisk. Consequently, a codon normally encoding an amino acid becomes a stop codon. \newline How will this alteration influence DNA $<$question\_type$>$? \newline Model Answer: \newline 1. Correct: $<$sample\_correct$>$ \newline 2. Incomplete/Irrelevant: $<$sample\_incomplete$>$ \newline 3. Incorrect: $<$sample\_incorrect$>$. \newline Student Answer: \newline
1. $<$substring$>$ \newline Which of the following categories does the student response fall in? \newline
A. Correct \newline
B. Incorrect \newline
C. Incomplete
}

These experiments were performed with three different temperatures (0, 0.5, 1), which is an OpenAI parameter that determines the creativity of the model. The usual range is from 0 to 2 with higher values making the output more random, while lower values making it more focused and deterministic. When the temperature is set to 0, the model uses log probability to automatically increase the temperature until certain thresholds are hit. We also performed experiments with higher temperatures like 1.5 and 2.0, but the output from the model was gibberish in most cases, so we do not report any results from these experiments.

\begin{table}[h]
\centering
\caption{Performance metrics of various BERT models as evaluated on the corresponding testing dataset, namely $T'_1$ for $BMQ_1$, $T'_2$ for $BMQ_2$ and $BMQ_1Q_2$, and $T'_3$ for $BMQ_3$ and $BMQ_1Q_2Q_3$}
\begin{tabular}{l>{\centering\arraybackslash}p{1cm}>{\centering\arraybackslash}p{1.3cm}>{\centering\arraybackslash}p{1cm}>{\centering\arraybackslash}p{1cm}>{\centering\arraybackslash}p{1cm}>{\centering\arraybackslash}p{1.3cm}>{\centering\arraybackslash}p{1cm}}
\hline
\textbf{Model Name} & \textbf{Mean Acc.} & \textbf{Median Acc.} & \textbf{Std. Dev. (SD)} & \textbf{Max Acc.} & \textbf{\% data for max} & \textbf{Acc. within 1 SD} & \textbf{\% data for 1 SD}\\
\hline
$BMQ_1Q_2$ & 89.48 & 90.32 & 05.05 & 91.42 & 82.5 & 91.32 & 62.5 \\
\hline
$BMQ_1Q_2Q_3$ & 82.81 & 84.23 & 09.96 & 86.93 & 82.5 & 86.93 & 82.5 \\
\hline
\end{tabular}
\label{table:chosenmodels}
\end{table}

\begin{table}[!h]
\caption{Examples of student responses and their classification across the three questions, with each response shown given by a different student. These sample responses have been used in the GPT prompts.} 
\centering
\begin{tabular}{>{\raggedright\arraybackslash}m{2.3cm}>{\raggedright\arraybackslash}m{3.5cm}>{\raggedright\arraybackslash}m{4.1cm}>{\raggedright\arraybackslash}m{3.7cm}}
\hline
\makecell{Question} & \makecell{\textless sample\_correct\textgreater} & \makecell{\textless sample\_incomplete\textgreater} & \makecell{\textless sample\_incorrect\textgreater} \\
\hline
\makecell{Replication} & 
It will not have any effect on the replication process. & 
This would be an example of a nonsense mutation. & 
The DNA will stop replicating when it reaches the stop codon. \\
\hline
\makecell{Transcription} & 
It will not have any effect on transcription. & 
This will cause a mutation in the transcription process. & 
In the process of transcribing DNA into RNA, the newly added stop codon will inhibit the rest of the chain from being transcribed into RNA. \\
\hline
\makecell{Translation} & 
This will influence translation because the stop codon will cause the amino acid sequence to end before it should. This will create a different polypeptide or protein that will either not function or function differently than it should have. & 
The code will be translated with a different base and will be read differently. This will result in a different protein being built. & 
This will have no influence on translation. \\
\hline
\end{tabular}
\label{table:sampleresponse}
\end{table}

\begin{figure}[!h]
    \centering
    \includegraphics[width=0.6\columnwidth]{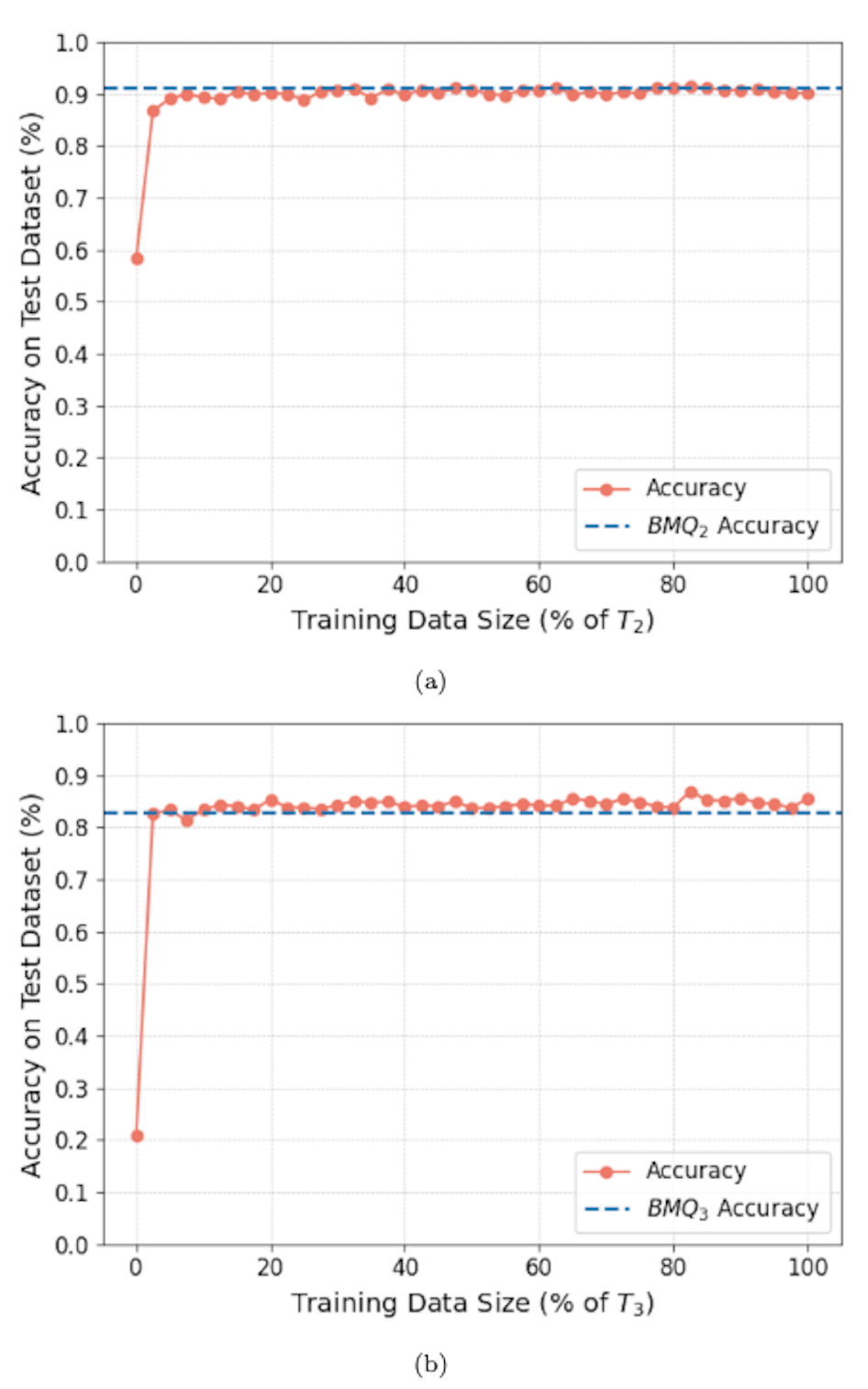}
    \caption{Advantages from Sequential Fine-Tuning with Near-Domain Data: Accuracy versus training data percentage for sequentially fine-tuned $BMQ_1...Q_j$ (j = 2, 3) models on transcription and translation questions. In particular, (a) Accuracy of $BMQ_1Q_2$ on the transcription task as training data increases shown in comparison to the accuracy of $BMQ_2$ and (b) Accuracy of $BMQ_1Q_2Q_3$ on the translation task as training data increases shown in comparison to the accuracy of $BMQ_3$. The results demonstrate that sequential fine-tuning on near-domain data enhances model performance on both tasks, achieving higher accuracy with less training data compared to $BMQ_i$ models. The results also show that the accuracy plateaus off as more data is fed into the model.}
    \label{FIG:BMQPTQData}
\end{figure}

\subsection{Declaration of generative AI and AI-assisted technologies in the writing process}
During the preparation of this work, the author(s) used popular large language models (LLMs), including Claude Sonnet by Anthropic and GPT models by OpenAI, solely to enhance the clarity and structure of the text. After using these tools, the author(s) thoroughly reviewed and edited the content to ensure accuracy and originality, and take(s) full responsibility for the content of the published article.

\section{Results}
We share the main experiments and their key results in this section.
\subsection{Base BERT on $Q_1$, $Q_2$, $Q_3$}
Starting from Base BERT $B_0$, and fine-tuning on entirety of the training dataset for $Q_1$, $Q_2$, and $Q_3$, we get $BMQ_1$, $BMQ_2$, and $BMQ_3$ which have an accuracy of 90.72\%, 91.22\%, and 82.83\% respectively. Without any training, $B_0$ gives an accuracy of 16.87\% on $T'_1$ (test data of $Q_1$), 14.77\% on $T'_2$ (test data of $Q_2$), and $B_0$ gives an accuracy of 13.37\% on $T'_3$ (test data of $Q_3$). This observed difference in model performance highlights the importance of fine-tuning, as the base model trained on general data does not generalize well to specialized domains such as biology without further adaptation. More metrics can be found in table \ref{table:initialdata} and table \ref{table:allmetrics}.

\subsection{Pretrained model results on $Q_1$, $Q_2$, $Q_3$}
As a reminder, models $BMQ_i$ and $BMQ_1...Q_j$ (j = 2, 3) differ from each other in the method used to train them. Model $BMQ_i$ is trained directly on Base BERT using the question $i$ whereas model $BMQ_1Q_2$ is $BMQ_1$ trained on subset of $T_2$ and model $BMQ_1Q_2Q_3$ is $BMQ_1Q_2$ trained on subset of $T_3$. For Question 2, $BMQ_1Q_2$ achieves an accuracy of 91.32\% after training on 62.5\% of the data which is the accuracy within 1 standard deviation of the average accuracy of the top 5 models and uses the least amount of training data. We see that the model accuracy for $BMQ_1Q_2$ started at 58.38\% (Figure \ref{FIG:BMQPTQData}). However, the accuracy eventually plateaus off i.e., the accuracies stabilize after a given point and show only negligible point differences later on. This suggests that near-domain data from other questions is helpful in improving accuracy on a particular question but only to a particular extent. For Question 3, $BMQ_1Q_2Q_3$ achieves an accuracy of 86.93\% after training on 82.5\% of the data, which is the accuracy within 1 standard deviation of the average accuracy of the top 5 models and uses the least amount of training data. This time, $BMQ_1Q_2Q_3$ starts with a very low accuracy initially, precisely 20.86\%. However, after training on merely 2.5\% of $T'3$, the accuracy jumps to 82.63\% and it also eventually plareaus off on adding more data. This suggests that $BMQ_1Q_2Q_3$ was able to learn insights from $Q_1$ and $Q_2$ that helped improve accuracy on $Q_3$ by just utilizing a small amount of data. More metrics can be found in table \ref{table:initialdata} and table \ref{table:allmetrics}.

\begin{table}[h]
\centering
\caption{Comparison of model accuracy between fine-tuned $B_0$ and sequentially fine-tuned models ($BMQ_1...Q_j$ (j = 2, 3)) under different data conditions, namely accuracy on training with 0\% data and the percentage of data required to achieve the comparable baseline accuracy for Transcription ($Q_2$) and Translation ($Q_3$) questions}
\label{table:initialdata}
\begin{tabular}{l>{\raggedright\arraybackslash}p{6.5cm}c}
\hline
\textbf{Question} & \textbf{Percentage of data used} & {\centering\textbf{$BMQ_1...Q_j$ (j = 2, 3)}}\\
\hline
\multirow{2}{*}{Transcription}
 & Accuracy when trained on 0\% data & 58.38 \\
 & Percentage of data needed to achieve comparable baseline accuracy & 7.5\% \\
 \hline
\multirow{2}{*}{Translation}
 & Accuracy when trained on 0\% data & 66.37 \\
 & Percentage of data needed to achieve comparable baseline accuracy & 2.5\% \\
 \hline
\end{tabular}
\end{table}

\subsection{Non-fine-tuned GPT Results} 

We use the same prompt as described in section \ref{sub:2.3.2} to prompt GPT 3.5 Turbo, GPT 4 Turbo (gpt-4-0125-preview endpoint), and GPT 4o. We perform these experiments with different temperatures as described in section \ref{sub:2.3.2}. For each of the GPT model experiments, the results obtained from using different temperatures are very close to each other as seen in Table \ref{tab:model_comparison}. Thus, we can consider the different accuracy from changing temperatures to be insignificant and thus, we feel comfortable comparing results for each model using the mean accuracy. Across all the three questions, GPT 3.5 Turbo does not perform very well with a gap of 5\% - 10\% from the best performing model out of the non-fine tuned versions of GPT in each category. This gap is maximum for $Q_3$ (9.78\%), lesser for $Q_2$ (9.08\%), and the least for $Q_1$ (8.05\%). Comparing GPT 4 Turbo and GPT 4o, the former performs way better than the latter for $Q_2$ and $Q_3$, surpassing the accuracy by 7.29\% and 10.05\% on those questions respectively. However, GPT 4o beats GPT 4 Turbo on $Q_1$ by a small margin of 1.23\%. It is interesting to note that for $Q_3$, GPT 3.5 Turbo also surpasses the performance of GPT 4o, although by merely 0.27 points.

We do not have a good reasoning on why some architectures perform better on some questions than others. Just like Shazeer mentions in his work on the GLU variants, "we attribute their success, as all else, to divine benevolence," referring to the somewhat unpredictable and often unexplained nature of certain architectures' performance despite seemingly similar conditions and hyperparameters.\citep{Shazeer2020GLUVariants}

\subsection{Fine-tuned GPT Results}
Using the same prompt and multiple temperatures as described in section \ref{sub:2.3.2}, we fine-tune GPT 3.5 Turbo (the only model that OpenAI allowed to fine-tune at the time these experiments were performed). On $Q_1$, we get an accuracy of 93.71\%, 93.81\%, 93.41\% with temperatures 0, 0.5, and 1 respectively. On $Q_2$, we get an accuracy of 93.91\% with all temperatures 0, 0.5, and 1. On $Q_3$, we get an accuracy of 88.02\%, 87.62\%, 87.92\% with temperatures 0, 0.5, and 1 respectively. The rest of the metrics can be found in table \ref{table:allmetrics}.

\renewcommand{\arraystretch}{1.2} 

\begin{longtable}{@{}p{4.2cm}p{2cm}cccc@{}}
\caption{Comprehensive comparison of the metrics (macro versions, where applicable) used to evaluate the performance of the different BERT and LLM models on their respective questions. N/A means not-applicable, indicating that no training happened in the experimental setting.}
\label{table:allmetrics} \\

\toprule
\textbf{Model Name} & \textbf{Training Data \%age} & \textbf{Accuracy} & \textbf{Precision} & \textbf{Recall} & \textbf{F1} \\
\midrule
\endfirsthead

\toprule
\textbf{Model Name} & \textbf{Training Data \%age} & \textbf{Accuracy} & \textbf{Precision} & \textbf{Recall} & \textbf{F1} \\
\midrule
\endhead

\midrule
\multicolumn{6}{r}{\small\itshape Continued on next page...} \\
\midrule
\endfoot

\bottomrule
\endlastfoot


$BMQ_1$ & 100 & 90.72 & 86.93 & 88.40 & 87.58 \\
gpt-3.5-turbo ($Q_1$, fine-tuned, temperature=0) & 100 & 93.71 & 91.51 & 91.00 & 91.20 \\
gpt-3.5-turbo ($Q_1$, fine-tuned, temperature=0.5) & 100 & 93.81 & 91.69 & 91.06 & 91.32 \\
gpt-3.5-turbo ($Q_1$, fine-tuned, temperature=1) & 100 & 93.41 & 91.10 & 90.54 & 90.76 \\
gpt-3.5-turbo ($Q_1$, non-fine-tuned, temperature=0) & N/A & 69.46 & 65.77 & 68.14 & 64.36 \\
gpt-3.5-turbo ($Q_1$, non-fine-tuned, temperature=0.5) & N/A & 68.76 & 65.34 & 67.47 & 63.69 \\
gpt-3.5-turbo ($Q_1$, non-fine-tuned, temperature=1) & N/A & 66.37 & 63.67 & 65.39 & 61.73 \\
gpt-4-0125-preview ($Q_1$, non-fine-tuned, temperature=0) & N/A & 74.75 & 71.24 & 65.45 & 61.71 \\
gpt-4-0125-preview ($Q_1$, non-fine-tuned, temperature=0.5) & N/A & 75.25 & 70.94 & 66.03 & 62.47 \\
gpt-4-0125-preview ($Q_1$, non-fine-tuned, temperature=1) & N/A & 75.05 & 69.56 & 65.49 & 61.65 \\
gpt-4o ($Q_1$, non-fine-tuned, temperature=0) & N/A & 77.15 & 70.99 & 73.96 & 71.03 \\
gpt-4o ($Q_1$, non-fine-tuned, temperature=0.5) & N/A & 76.95 & 71.23 & 74.41 & 71.34 \\
gpt-4o ($Q_1$, non-fine-tuned, temperature=1) & N/A & 74.65 & 69.17 & 72.20 & 68.87 \\
\midrule
$BMQ_2$ & 100 & 91.22 & 87.70 & 86.34 & 86.98 \\
$BMQ_1Q_2$ & 62.5 & 91.32 & 88.14 & 86.09 & 86.87 \\
gpt-3.5-turbo ($Q_2$, fine-tuned, temperature=0) & 100 & 93.91 & 90.43 & 92.95 & 91.53 \\
gpt-3.5-turbo ($Q_2$, fine-tuned, temperature=0.5) & 100 & 93.91 & 90.43 & 92.95 & 91.53 \\
gpt-3.5-turbo ($Q_2$, fine-tuned, temperature=1) & 100 & 93.91 & 90.43 & 92.95 & 91.53 \\
gpt-3.5-turbo ($Q_2$, non-fine-tuned, temperature=0) & N/A & 64.77 & 58.07 & 58.39 & 57.28 \\
gpt-3.5-turbo ($Q_2$, non-fine-tuned, temperature=0.5) & N/A & 63.37 & 56.86 & 56.64 & 55.71 \\
gpt-3.5-turbo ($Q_2$, non-fine-tuned, temperature=1) & N/A & 63.97 & 57.08 & 57.63 & 56.44 \\
gpt-4-0125-preview ($Q_2$, non-fine-tuned, temperature=0) & N/A & 73.05 & 58.46 & 60.38 & 55.24 \\
gpt-4-0125-preview ($Q_2$, non-fine-tuned, temperature=0.5) & N/A & 73.25 & 59.16 & 60.62 & 55.67 \\
gpt-4-0125-preview ($Q_2$, non-fine-tuned, temperature=1) & N/A & 73.05 & 58.77 & 60.50 & 55.56 \\
gpt-4o ($Q_2$, non-fine-tuned, temperature=0) & N/A & 67.56 & 64.99 & 65.71 & 62.50 \\
gpt-4o ($Q_2$, non-fine-tuned, temperature=0.5) & N/A & 65.77 & 64.17 & 63.76 & 60.93 \\
gpt-4o ($Q_2$, non-fine-tuned, temperature=1) & N/A & 64.17 & 62.21 & 61.12 & 58.85 \\
\midrule
$BMQ_3$ & 100 & 82.83 & 72.88 & 69.84 & 70.70 \\
$BMQ_1Q_2Q_3$ & 82.5 & 86.93 & 79.91 & 74.65 & 76.70 \\
gpt-3.5-turbo ($Q_3$, fine-tuned, temperature=0) & 100 & 88.02 & 82.60 & 75.80 & 78.22 \\
gpt-3.5-turbo ($Q_3$, fine-tuned, temperature=0.5) & 100 & 87.62 & 81.83 & 74.97 & 77.39 \\
gpt-3.5-turbo ($Q_3$, fine-tuned, temperature=1) & 100 & 87.92 & 82.37 & 75.55 & 77.94 \\
gpt-3.5-turbo ($Q_3$, non-fine-tuned, temperature=0) & N/A & 68.56 & 61.82 & 56.89 & 56.99 \\
gpt-3.5-turbo ($Q_3$, non-fine-tuned, temperature=0.5) & N/A & 68.46 & 61.94 & 57.04 & 57.15 \\
gpt-3.5-turbo ($Q_3$, non-fine-tuned, temperature=1) & N/A & 66.97 & 55.61 & 55.32 & 54.80 \\
gpt-4-0125-preview ($Q_3$, non-fine-tuned, temperature=0) & N/A & 77.84 & 66.09 & 66.12 & 66.10 \\
gpt-4-0125-preview ($Q_3$, non-fine-tuned, temperature=0.5) & N/A & 77.45 & 65.57 & 65.12 & 65.33 \\
gpt-4-0125-preview ($Q_3$, non-fine-tuned, temperature=1) & N/A & 78.04 & 66.99 & 66.87 & 66.92 \\
gpt-4o ($Q_3$, non-fine-tuned, temperature=0) & N/A & 67.56 & 68.58 & 62.71 & 60.66 \\
gpt-4o ($Q_3$, non-fine-tuned, temperature=0.5) & N/A & 67.37 & 67.20 & 62.30 & 60.05 \\
gpt-4o ($Q_3$, non-fine-tuned, temperature=1) & N/A & 68.26 & 66.62 & 62.67 & 60.27 \\

\end{longtable}

\section{Discussion}
\subsection{Sources of Error in Model Results}
\label{dis:zipf}
Upon evaluation of models on test data, we found several human-machine disagreements. In particular, as shown in Table \ref{table:humanmisscores}, $BMQ_1$ had 70 (in $T'_1$), $BMQ_2$ had 85 (in $T'_2$), $BMQ_3$ had 162 (in $T'_3$), $BMQ_1Q_2$ had 87 (in $T'_2$), and $BMQ_1Q_2Q_3$ had 131 (in $T'_3$) disagreements. With help from domain experts, we studied the different errors in the model predictions and classified them into three categories that are useful for further research in this field.

\begin{enumerate}
    \item \textbf{\emph{Human Errors}}: While evaluating the results given by the models, a domain expert re-evaluated the assigned classifications on miscoded responses. After this review, the domain expert also disagreed with the initial assigned score on about 10\% misscored cases (Table \ref{table:humanmisscores}), suggesting these responses may have mistakenly been given the wrong scores initially or were difficult for even experts to agree on classification. It's worth noting that the initial inter-grader agreement was around 80\%, indicating a fairly strong but not absolutely perfect baseline consistency among annotators. This further reinforces the idea that some level of ambiguity or subjectivity was present in the data from the beginning. This is quite exciting because this suggests that the model is able to learn well despite the presence of potential noise in the data. Upon second evaluation, $BMQ_1$ had 10, $BMQ_1$ had 8, $BMQ_3$ had 17, $BMQ_1Q_2$ had 12, and $BMQ_1Q_2Q_3$ had 11 possible human misscores.

\begin{table}[h]
\centering
\caption{Model performance comparison across three questions calculated from experiments with different temperatures as presented in table \ref{table:allmetrics}}
\begin{tabular}{p{2.3cm}p{3cm}>{\centering\arraybackslash}p{4cm}>{\centering\arraybackslash}p{4cm}}
\hline
\textbf{Question} & \textbf{Model Name} & \textbf{Mean Accuracy} & \textbf{Sample Standard Deviation} \\
\hline
\multirow{3}{*}{Replication}
 & GPT 3.5 Turbo (fine-tuned) & 93.64 & 0.21\\
 & GPT 3.5 Turbo (non-fine-tuned) & 68.20 & 1.62\\
 & GPT 4 Turbo (non-fine-tuned) & 75.02 & 0.25\\
 & GPT 4o (non-fine-tuned) & 76.25 & 1.39\\
\hline
\multirow{3}{*}{Transcription}
 & GPT 3.5 Turbo (fine-tuned) & 93.91 & 0.00\\
 & GPT 3.5 Turbo (non-fine-tuned) & 64.04 & 0.70\\
 & GPT 4 Turbo (non-fine-tuned) & 73.12 & 0.12\\
 & GPT 4o (non-fine-tuned) & 65.83 & 1.70\\
\hline
\multirow{3}{*}{Translation}
 & GPT 3.5 Turbo (fine-tuned) & 87.85 & 0.21\\
 & GPT 3.5 Turbo (non-fine-tuned) & 68.00 & 0.89\\
 & GPT 4 Turbo (non-fine-tuned) & 77.78 & 0.30\\
 & GPT 4o (non-fine-tuned) & 67.73 & 0.47\\
\hline
\end{tabular}
\label{tab:model_comparison}
\end{table}

\begin{table}[h]
\centering
\caption{Human errors: Number of samples that were counted as misclassified by the models, but were actually misunderstood because of original human miscodes.}
\label{table:humanmisscores}
\begin{tabular}{p{2.3cm}p{2.5cm}>{\centering\arraybackslash}p{3cm}>{\centering\arraybackslash}p{3cm}}
\hline
\textbf{Question} & \textbf{Model} & \parbox{2cm}{\centering \textbf{Human-machine disagreements}} & \parbox{2cm}{\centering\textbf{Possible Human Miscodes}}\\
\hline
\multirow{1}{*}{Replication}
 & $BMQ_1$ & 93 & 10 \\ 
\hline
\multirow{2}{*}{Transcription}
 & $BMQ_2$ & 88 & 8 \\ 
 & $BMQ_1Q_2$ & 87 & 12 \\
 \hline
\multirow{2}{*}{Translation}
 & $BMQ_3$ & 172 & 17 \\ 
 & $BMQ_1Q_2Q_3$ & 131 & 11 \\
 \hline
\end{tabular}
\end{table}

\item \textbf{\emph{Shared errors across models}}: For $Q_2$ and $Q_3$, we analyzed how many responses were not classified correctly by any of the $BMQ_i$ or $BMQ_1...Q_j$ (j = 2, 3) models i.e., which $Q_2$ responses were not classified correctly by both $BMQ_2$ and $BMQ_1Q_2$ and which $Q_3$ responses were not classified correctly by both $BMQ_3$ and $BMQ_1Q_2Q_3$. For each of the three human labels, there are two possible different classifications generated by $BMQ_i$ model and two possible different classifications generated by $BMQ_1...Q_j$ (j = 2, 3) model, which means there are twelve scenarios. In total, there are 62 such $Q_2$ responses and 102 such $Q_3$ responses (number of samples for each combination is shown in Table \ref{table:responseconstructionerror}).

    In $Q_2$, errors are more prominent in responses that are labeled "Incomplete" by human graders whereas in $Q_3$, errors are more prominent in responses that are labeled "Incorrect" by human graders. Specifically, the scenario “Incomplete, Incorrect, Incorrect” appears most often in $Q_2$ (17 instances), and “Incorrect, Incomplete, Incomplete” occurs most frequently in $Q_3$ (27 instances), highlighting the prevalence of wrong predictions whenever the human label or ground truth is not strictly “Correct.” Conversely, when the human label is “Correct,” relatively fewer instances fall into incorrect or incomplete predictions, implying that correct responses tend to be correctly identified by both models. We believe this pattern arises because the correct answer to a question is usually singular or limited, while an incorrect or incomplete answer can take on a variety of forms. Consequently, even if the model learns how to classify the correct response, accounting for every possible way in which an answer might be wrong or incomplete proves much more challenging.
    
    If we include the results from GPT models too, there are 14 responses in $Q_1$, 17 responses in $Q_2$ and 35 responses in $Q_3$ that are not classified correctly by any model. We attribute this behavior to the general heavy-tailed zipf distribution of student responses in natural language which has been noted by other works \citep{nguyen2014codewebs, piech2016uncovering}. A distribution is said to follow zipf's law if the probability of the $n$-th most probable outcome follows an inverse power-law with respect to $n$. This is to say that the probability of an outcome $o$ is proportional to:
    \begin{equation}
    p(o) \propto \frac{1}{\text{rank}(o)^{\beta}}
    \end{equation}
    where rank($o$) is calculated after ordering all the outcomes based on their probabilities, and $\beta$ is the exponent parameter guiding the shape of the distribution. Such distributions are called Zipfian distributions or power-law distributions. Useful inferences can be made from identifying the zipf nature of distributions, but the most relevant one is identifying that zipf distributions have a long tail which implies that a huge portion of the responses will only be observed exactly once and hence, it is difficult to generalize algorithms/models to accurately classify them because during test time, there will be new examples that can introduce new student strategies or unseen tokens or new misconceptions.
    
    This challenge is further compounded in natural language processing (NLP), where even two individuals attempting to express the same idea may do so using different lexical or syntactic constructions. That is, semantically similar responses can vary significantly in wording, leading to a proliferation of unique combinations of tokens. Consequently, even correct or meaningful responses may appear rare or entirely novel to a model simply due to surface-level differences which increase the difficulty of building robust systems that can generalize beyond frequently seen patterns. As discussed by \citet{Piantadosi2014Zipf}, Zipfian distributions are a fundamental characteristic of human language, reflecting a balance between communicative efficiency and cognitive constraints. In educational settings and open-response evaluation tasks, this implies that a small number of responses or linguistic patterns will be very frequent, while the majority will be rare or even unique. Models must therefore be especially sensitive to these low-frequency, long-tail responses—those that may only appear once but nonetheless represent meaningful, valid, or insightful contributions. Ignoring this Zipfian behavior risks systematic underperformance on responses that differ from the norm in form but not in substance, including novel misconceptions or creative solution strategies that are pedagogically significant.

\begin{table}[h]
\centering
\caption{Shared errors across models: Number of samples where both models predicted wrong output (for $Q_2$ and $Q_3$)}
\begin{tabular}{p{2.4cm}lcccc}
\hline
\textbf{Question} & \textbf{Human Label} & \textbf{$BMQ_i$} & \textbf{$BMQ_1...Q_j$ (j = 2, 3)} & \textbf{Total}\\
\hline
\multirow{12}{*}{\begin{tabular}[c]{@{}c@{}}Transcription\\($Q_2$)\end{tabular}}
 & Correct & Incomplete & Incomplete & 14 \\
 & Correct & Incomplete & Incorrect & 0 \\
 & Correct & Incorrect & Incomplete & 1 \\
 & Correct & Incorrect & Incorrect & 4 \\
 & Incomplete & Correct & Correct & 12 \\
 & Incomplete & Correct & Incorrect & 3 \\
 & Incomplete & Incorrect & Correct & 1 \\
 & Incomplete & Incorrect & Incorrect & 17 \\
 & Incorrect & Correct & Correct & 3\\
 & Incorrect & Correct & Incomplete & 0 \\
 & Incorrect & Incomplete & Correct & 2 \\
 & Incorrect & Incomplete & Incomplete & 5 \\
\hline
\multirow{12}{*}{\begin{tabular}[c]{@{}c@{}}Translation\\($Q_3$)\end{tabular}}
 & Correct & Incomplete & Incomplete & 7 \\
 & Correct & Incomplete & Incorrect & 1 \\
 & Correct & Incorrect & Incomplete & 1 \\
 & Correct & Incorrect & Incorrect & 2 \\
 & Incomplete & Correct & Correct & 17 \\
 & Incomplete & Correct & Incorrect & 5 \\
 & Incomplete & Incorrect & Correct & 2 \\
 & Incomplete & Incorrect & Incorrect & 7 \\
 & Incorrect & Correct & Correct & 26 \\
 & Incorrect & Correct & Incomplete & 2 \\
 & Incorrect & Incomplete & Correct & 5 \\
 & Incorrect & Incomplete & Incomplete & 27 \\
\hline
\end{tabular}
\label{table:responseconstructionerror}
\end{table}

    \item \textbf{\emph{Model-specific errors}}: For both $Q_2$ and $Q_3$, we also study the responses that were classified incorrectly by one model, but were classified correctly by the other. For each of the three human labels, we are seeing which of the two models is providing a different classification, and there are two ways in which the classification can be different. Thus, there are twelve total scenarios, and the number of samples across them are listed in Table \ref{table:modelspecific}. From the data, we see that most misclassifications involve the “Incomplete” category—either the model predicts “Incomplete” when the ground truth is “Correct” or “Incorrect,” or it predicts “Correct”/“Incorrect” where the ground truth is “Incomplete.” This pattern is somewhat expected, as “Incomplete” often represents a middle ground between “Correct” and “Incorrect,” making it more susceptible to boundary confusion during classification. For example, this is most noticeable in $Q_3$ where there are 14 scenarios of "Incomplete" response predicted as "Correct" by $BMQ_3$, 17 scenarios of "Incorrect" response predicted as "Incomplete" by $BMQ_3$, and 12 scenarios of "Incorrect" response being predicted as "Incorrect" by $BMQ_1Q_2Q_3$, which outnumbers the direct "Correct" to "Incorrect" or vice versa swaps. In other words, the biggest share of errors stems from confusion involving the “Incomplete” label, rather than straightforward flips between “Correct” and “Incorrect.” The scenarios where either of the models predicts the correct class as incorrect or vice versa are limited.
    
    Based on domain expert evaluation, there were some themes that were noticed across these models. For the Transcription question ($Q_2$), many of the responses misclassified only by $BMQ_2$ (and correctly coded by $BMQ_1Q_2$) involved content that is actually relevant to the subsequent Translation question ($Q_3$) rather than strictly to Transcription. Moreover, $BMQ_2$ produced more unique misclassifications on responses that compared an abnormal transcription process (arising from the mutation) with a normal process. For the Translation question ($Q_3$), the $BMQ_1Q_2Q_3$ model made more errors than $BMQ_3$ on responses containing detailed descriptions of the translation steps or definitions of different types of mutations. Conversely, $BMQ_3$ had more unique misclassifications on responses discussing the consequences of cells possessing mutated proteins.

\begin{table}[h]
\centering
\caption{Model-specific errors: Number of samples where one model predicted wrong output whereas the other model predicted correct output (for $Q_2$ and $Q_3$)}
\begin{tabular}{p{2.4cm}lcccc}
\hline
\textbf{Question} & \textbf{Human Label} & \textbf{$BMQ_i$} & \textbf{$BMQ_1...Q_j$ (j = 2, 3)} & \textbf{Total}\\
\hline
\multirow{12}{*}{\begin{tabular}[c]{@{}c@{}}Transcription\\($Q_2$)\end{tabular}}
 & Correct & Correct & Incomplete & 3 \\
 & Correct & Correct & Incorrect & 5 \\
 & Correct & Incomplete & Correct & 5 \\
 & Correct & Incorrect & Correct & 2 \\
 & Incomplete & Correct & Incomplete & 5 \\
 & Incomplete & Incomplete & Correct & 8 \\
 & Incomplete & Incomplete & Incorrect & 7 \\
 & Incomplete & Incorrect & Incomplete & 5 \\
 & Incorrect & Correct & Incorrect & 2 \\
 & Incorrect & Incomplete & Incorrect & 4 \\
 & Incorrect & Incorrect & Correct & 1 \\
 & Incorrect & Incorrect & Incomplete & 1 \\
\hline
\multirow{12}{*}{\begin{tabular}[c]{@{}c@{}}Translation\\($Q_3$)\end{tabular}}
 & Correct & Correct & Incomplete & 3 \\
 & Correct & Correct & Incorrect & 2 \\
 & Correct & Incomplete & Correct & 7 \\
 & Correct & Incorrect & Correct & 7 \\
 & Incomplete & Correct & Incomplete & 14 \\
 & Incomplete & Incomplete & Correct & 5 \\
 & Incomplete & Incomplete & Incorrect & 12 \\
 & Incomplete & Incorrect & Incomplete & 9 \\
 & Incorrect & Correct & Incorrect & 6 \\
 & Incorrect & Incomplete & Incorrect & 17 \\
 & Incorrect & Incorrect & Correct & 4 \\
 & Incorrect & Incorrect & Incomplete & 3 \\
 
\hline
\end{tabular}
\label{table:modelspecific}
\end{table}

\end{enumerate}

\subsection{Advantages of Employing Near-Domain Data}
As can be seen in Figure \ref{FIG:BMQPTQData} and Table \ref{table:allmetrics}, for both $Q_2$ and $Q_3$, the $BMQ_1...Q_j$ (j = 2, 3) models achieve comparable or even better performance than $BMQ_i$ models using lesser amount of data. The performance of these $BMQ_1...Q_j$ (j = 2, 3) models level off as more data is fed into them. However, we tried to better understand the efficacy of using near-domain data and observed that using near-domain data offers two key advantages: \textit{data advantage} and \textit{accuracy advantage}.

Data Advantage refers to the ability of $BMQ_1...Q_j$ (j = 2, 3) models trained on near-domain data to achieve the same level of accuracy as the $BMQ_i$ (i = 1, 2, 3) model trained on 100\% of the target data, but with significantly less training data.  We observed this pattern for both questions. For $Q_2$, $BMQ_1Q_2$ achieves the same accuracy as $BMQ_2$ using only 62.5\% of the training data (Table \ref{table:allmetrics}). In fact, $BMQ_1Q_2$ reaches ~90\% accuracy with just 7.5\% of the data, after which its performance plateaus around 90\%–92\% (Figure \ref{FIG:BMQPTQData}). For $Q_3$, $BMQ_1Q_2Q_3$ matches the performance of $BMQ_3$ with only 2.5\% of the training data. These findings suggest that prior exposure to related concepts through near-domain data enables the model to generalize more effectively, thereby significantly reducing the need for extensive manual annotation.

Our approach is a form of few-shot learning, where the model generalizes to a new question with only a small number of labeled examples by leveraging prior training on conceptually related (near-domain) questions. Rather than learning from scratch, the $BMQ_1...Q_j$ (j = 2, 3) models benefit from shared conceptual structure across questions (e.g., within the Central Dogma), allowing them to generalize more quickly and accurately with minimal labeled data from the new target question. Therefore, our approach represents a domain-adaptive, few-shot learning idea as a practical and scalable solution for real-world educational settings, where collecting large amounts of labeled data for every new assessment is infeasible.

This means that instructors in academic settings can achieve the same automated grading accuracy with significantly less labeled data if they have access to near-domain data. A data advantage implies that a model trained on responses to related questions requires fewer labeled examples to match the performance of a model trained solely on the target question. This greatly reduces manual grading effort, making automated grading more efficient and scalable in real-world classrooms. Instructors can leverage responses from previously asked, related questions—especially on the same topic—to compensate for limited labeled data on new questions. Instead of grading hundreds of responses for every new question, they only need to label a small subset, enabling the model to generalize from past knowledge and streamline the grading process.

Accuracy Advantage, on the other hand, refers to the model’s ability to achieve higher accuracy with the same amount of training data when it has been exposed to related, near-domain data. This occurs when $BMQ_1...Q_j$ (j = 2, 3) models yield higher accuracy than the $BMQ_i$ model using the same amount of data. The advantage was evident when comparing the performance of $B_0$ on $Q_2$ (14.77\%) and $Q_3$ (13.37\%) with $BMQ_1Q_2$ (58.38\%) and $BMQ_1Q_2Q_3$ (20.83\%), where all models were trained on 0\% of the target data. In addition, compared to $BMQ_3$, the $BMQ_1Q_2Q_3$ model offers accuracy advantage when both are trained on the 100\% training dataset (Figure \ref{FIG:BMQPTQData}). In fact, one can argue that $BMQ_1Q_2Q_3$ offers both accuracy and data advantage, as it achieves a higher accuracy (86.93\%) with less training data (82.5\%) compared to $BMQ_3$ (82.83\%), which is trained on 100\% of the training dataset.

Even with a limited number of labeled responses for a new question, a model can achieve higher grading accuracy by leveraging knowledge from similar, near-domain questions. For example, if a model has already been trained to grade answers about DNA transcription, it can more effectively classify responses about DNA translation—even when the latter has only a small set of labeled examples. This accuracy advantage means that the model benefits from shared concepts, terminology, and response structures across related questions, allowing it to generalize better without requiring extensive new annotations.

This is particularly useful in academic settings where instructors often face constraints on time and resources for manual grading. Instead of labeling hundreds of responses for each new question, they can rely on models that have learned from previous, related questions to maintain grading quality. This not only reduces the likelihood of misclassification when labeled data is scarce but also ensures greater fairness and reliability—especially important in high-stakes educational environments. By using near-domain data, instructors can make automated grading more efficient, scalable, and practical for real-world classrooms.

We note that the accuracy advantage in $Q_3$ (improvement of 7.46\%) is relatively smaller compared to that in $Q_2$ (improvement of 43.61\%) when trained on 0\% data. We believe that the reason that we see a low initial accuracy on $Q_3$ is because unlike $Q_1$ and $Q_2$, where the correct answer is that there will be no effect on replication and transcription respectively, the correct answer for $Q_3$ is that there would be an effect on translation. Since the $BMQ_1Q_2Q_3$ model does not have any context about the change of question or change of correct answer, it did not perform very well initially. However, when trained with only a few trained examples (2.5\%), data advantage kicks in and we get high accuracy performance.

\subsection{Fine-tuned vs Non-Fine-Tuned GPT 3.5 Turbo}
From table \ref{tab:model_comparison}, we can clearly see that fine-tuning GPT 3.5 Turbo performs way better than its non-fine-tuned counterpart with a margin of 25.446\%, 29.873\%, and 19.856\% on $Q_1$, $Q_2$, and $Q_3$ respectively. This suggests that although GPT models are trained on vast amounts of diverse data, they struggle to generalize effectively to specific downstream tasks without targeted fine-tuning. Feeding information particular to the question under consideration helps them better understand the patterns. In other words, despite having seen a wide range of data, the non-fine-tuned models lack the necessary task-specific grounding, leading to poor performance on unseen, domain-specific examples. Providing fine-tuned data aligned with the structure and intent of each question allows the model to better capture relevant patterns and nuances.

\subsection{Overall Performance across Questions}
An interesting result to note is that the performance of both the fine-tuned GPT model and fine-tuned BERT models is consistently reduced for $Q_3$ compared to that of $Q_1$ and $Q_2$ across all metrics - accuracy, precision, recall, and F1 score. This trend is consistent with previous results for ASAG based on this question \citep{Prevost2016}. Examining these misclassifcations, we found that the most errors resulted from both approaches assigning the same code, but different from the human label (see Table \ref{table:humanmisscores}). Based on human experts, translation is considered to be a more difficult concept for students to master because developing a proper understanding of it requires knowledge of the preceding processes (e.g., transcription) and molecules (e.g., DNA and RNA). Consequently, even the models found it more challenging to evaluate responses to this question accurately. We also noted that a number of human disagreements occur assigning either Incomplete or Incorrect label to Translation responses. This may suggest that coding critera could be improved or there are many borderline responses between these categories. Previous work found that responses coded as Incorrect for Translation include a wide variety of ideas and concepts, more so than for responses to Replication and Transcription question, representing diverse, yet incorrect, thinking about the science \citep{Uhl2021}. 

\subsection{Model Size and Environment}

The results of this study suggest that the best performing models are the fine-tuned versions of BERT and GPT. Using the LLM average accuracy from Table \ref{tab:model_comparison} and the BERT derivative model accuracy from \ref{table:allmetrics}, we observed that fine-tuned GPT 3.5 Turbo surpasses the fine-tuned BERT model by 2.59\% and 0.92\% on $Q_2$, and $Q_3$ respectively. Fine-tuned GPT 3.5 Turbo surpasses $BMQ_1$ by 2.92\% as well. However, it is important to note that the GPT models are quite huge and consist of a large number of parameters (in the scale of hundreds of billions or trillions depending on the precise model) compared to BERT with lesser number of parameters (in the scale of hundreds of millions). In addition,  GPT models require significant training time. Fine-tuning and training large models demand significant time, energy, and computational resources, resulting in substantial carbon dioxide emissions that negatively impact the environment \citep{agarwal2024hidden}. While the performance gains may appear negligible in individual instances, when scaled across thousands or millions of grading tasks, the cumulative environmental and computational costs become far more consequential. In light of the growing climate-related concerns, this tradeoff in accuracy seems quite reasonable and suggests that training small models for automated grading tasks would be more appropriate than utilizing foundational and large language models like GPT.

\subsection{Significance of automated grading}
\subsubsection{Efficiency in grading}
\label{section:efficient_grading}
Providing feedback at the right time and frequency for various topics throughout the semester is crucial for fostering effective learning. Automated grading systems ensure rapid and consistent evaluation of student responses, thereby allowing instructors to focus their attention on pedagogy and student learning experience, rather than administrative grading tasks. By providing iterative learning opportunities through timely feedback, these systems allow students to reflect on their thought process, correct any misunderstandings and eventually, improve their performance progressively. Moreover, automated grading techniques enhance scalability, making frequent assessments feasible even in large or multi-disciplinary courses with low teacher-to-student ratio. With less time spent on manual evaluations, educators can devote more effort to teaching and mentoring. This allows instructors to devote more effort to curriculum and instructional design, enhancing the quality of learning materials. Furthermore, data from previous assessments can be leveraged to reshape future course content, addressing knowledge gaps, and tailoring instruction to student needs.

\subsubsection{Consistency and Fairness}
Consistency and fairness are critical aspects of any grading system, and automated grading techniques excel in addressing these concerns. Automation ensures uniform grading standards, eliminating human bias and variability that can arise from subjective judgment or fatigue.  While human grading often results in inconsistencies—as seen in the dataset used in our study where the human-labels have an intergrader agreement of 80\%—automated systems evaluate responses based on predefined criteria, leading to more objective and transparent assessments that ensure fair treatment for all students. Since the automated systems only see the information provided them in input, which is the student response, their outputs are irrespective of any other student information that could make a human grader biased. This balance of fairness and consistency in output strengthens the credibility of the grading process while promoting student growth.

\subsubsection{Self-assessment tool}
Automated grading systems support self-assessment, allowing students to gauge their understanding without the pressure of affecting their final grades. Since the students have the flexibility to interact ith these systems on their own time and pace, it also helps them prepare more effectively. By reducing test anxiety, they minimize its impact on performance, creating a more equitable learning experience. Additionally, self-assessment can uncover hidden expectations and areas requiring further preparation, enabling students to track their progress and see tangible evidence of their learning over time. When the data is shared with instructors, it can also help them learn better insights about student learning experience.

\section{Conclusion}
Providing automated feedback at scale is a challenging problem due to various factors, including student response variability, subjectivity in grading, and the computational demands of AI models. In this work, we proposed a novel framework that leverages near-domain data to enhance short-answer grading accuracy while requiring significantly less labeled data. By fine-tuning BERT-based models, we demonstrated that our approach outperforms state-of-the-art large language models (LLMs) like GPT-3.5, GPT-4, and GPT-4o, even when these models are provided with reference answers. The key takeaways from our work are the accuracy advantage, which refers to the model’s ability to achieve higher accuracy with the same amount of training data when exposed to related, near-domain data; and the data advantage, which captures the model’s ability to reach the same target accuracy as a model trained on 100\% of the target data, but using significantly less training data due to prior exposure to near-domain data. These findings have significant implications for scalable, resource-efficient automated grading systems. These results demonstrate that incorporating prior knowledge from related questions can enhance generalization, reduce annotation demands, and improve the scalability of automated feedback systems in educational contexts.

Beyond grading performance, our study also analyzed patterns of grading errors, identifying cases where AI models diverged from human-labeled ground truth. This analysis sheds light on key challenges in automated grading, particularly in ensuring fairness, explainability, and adaptability across different domains. Our comparative analysis of high-compute models (GPT-3.5, GPT-4, GPT-4o) versus low-compute models (BERT and fine-tuned versions) revealed important insights. While fine-tuning GPT models provides the best performance, this approach is often impractical due to high computational costs and energy demands. In contrast, fine-tuned BERT models provide a strong alternative, offering high accuracy without requiring the extensive computational power needed for large-scale LLMs.

While our framework presents a scalable and high-performing solution, it is not without limitations that could be removed in future. First, the dataset used in this study has an inter-grader agreement of 80\%, meaning there is some level of subjectivity in human-labeled data, which can impact model performance. Second, our study focuses on a specific topic (Central Dogma in Molecular Biology), making it difficult to generalize results to other subjects and disciplines. Future work should evaluate our approach across diverse domains to validate its broader applicability. Third, while we fine-tuned BERT, we did not have access to fine-tune GPT-4 or other large-scale models, which could provide additional insights into how different architectures benefit from near-domain data. Future work could explore evaluations with DeepSeek, Gemini, and other emerging LLMs, as well as fine-tuning experiments with GPT-4 Turbo and other proprietary models to further benchmark and validate our findings.

Despite these limitations, our work offers valuable contributions to the field of AI-driven education. The insights gained from near-domain data adaptation and model efficiency trade-offs can inform future research on integrating automated grading into adaptive learning environments. Additionally, the classifications produced by our framework could be extended to support formative feedback mechanisms, helping students not only receive scores but also understand their mistakes and improve their learning outcomes.

\subsection{Acknowledgements}
The authors would like to acknowledge the Center for Educational Effectiveness at UC Davis for providing feedback.

\section{Statements and Declarations}
\subsection{Funding}
The authors did not receive support from any organization for the submitted work.
\subsection{Competing Interests}
The authors declare that they have no competing interests.
\subsection{Ethics approval and consent to participate}
The authors have no competing interests to declare that are relevant to the content of this article.
\subsection{Consent for publication}
The manuscript does not contain any individual person's data in any form.
\subsection{Data availability}
The datasets used and/or analyzed during the current study are available from the corresponding author on reasonable request.
\subsection{Materials availability}
Any relevant materials regarding this study are available from the corresponding author on reasonable request.
\subsection{Code availability}
Any code files regarding this study are available from the corresponding author on reasonable request.
\subsection{Author contribution}
SA: Methodology; Software; Formal analysis; Visualization; Writing - Original Draft. AM: Conceptualization; Methodology; Supervision; Formal analysis; Resources; Writing - Review \& Editing. KCH: Data Curation; Writing - Review \& Editing.

\bibliography{sn-bibliography}

\end{document}